# A PRACTICAL APPROACH TO EXPRESSING DIGITALLY SIGNED DOCUMENTS

**Diana Berbecaru, Marius Marian**

*Dip. di Automatica e Informatica*
*Politecnico di Torino*
*Corso Duca degli Abruzzi24, 10129, Torino, Italy*

Abstract: Initially developed and considered for providing authentication and integrity functions, digital signatures are studied nowadays in relation to electronic documents (e-docs) so that they can be considered equivalent to handwritten signatures applied on paper documents. Nevertheless, a standardized format to be used specifically for e-doc representation was not yet specified. Each document management system is free to choose whatever e-doc format is suitable for its requirements (e.g. ASCII, Word, PDF, binary). So far, some solutions for document management systems were found but none of them was designed to consider security an important requirement and to enable digital signing and easy management of e-docs. A possible solution to this problem is our *secure* document management system named AIDA. This paper focuses on the use of XML and XML Signature for the representation of e-docs in AIDA.

## 1. INTRODUCTION

The main characteristic of the new electronic age is a shift from slow physical transportation of material goods to light-speed transmission of immaterial, but highly valuable, information. New communication systems, and especially Internet, could significantly modify the way in which public services are offered to citizens since they have new possibilities and an increasing access to the information highway.

However, expected change of the transactions and services are still shaded by concerns about *security* and legal uncertainty. E-mail systems, for example, and the documents they carry are susceptible to vulnerabilities such as leakage and modification of document contents or the impersonation of legitimate users. To be able to rely on the electronic records, assurance is needed that those records are authentic (i.e. that they really arrived from the person who claims to be the sender) and that they have not been altered. In most cases, confidentiality is required too. Moreover, in business transactions it is very important for a transaction to be not only authentic, with unaltered content, but also at the same time the sender must not be able to deny his action. This is generally known by the name of non-repudiation. In these conditions, security has become one of the major requirements in communications performed via Internet.

Public-key cryptography is nowadays the widely recognized technique to develop and deploy authentication, integrity, confidentiality and non-repudiation security services. For example, it is used to secure e-mail and web transactions, to create virtual private networks and to create legally binding digital signatures in several countries. Public-key cryptography was introduced in 1976 by Diffie & Hellman and makes use of a pair of mathematically related keys used for encryption and decryption; one key, named the private key is only known by the owner, while the other one, named the public key is publicly known. To securely bind a public key to an entity, a data structure named *public-key certificate* (PKC) or digital certificate is used.

One open issue in applying PKC in real world scenarios is the use of digital signatures in relation to e-docs so that they can be considered equivalent to handwritten signatures applied on paper documents. E-documents are not just an electronic transposition of a document but rather a data format that is secure, standard and is open (i.e. it can be read in 30 years because the specification is public).

The EC-funded AIDA project (IST-1999-10497) (AIDA, 1999) provides a secure document management architecture, which is based on standards and technologies available on the market today and cutting edge techniques, such as smart-cards and handheld PCs. The AIDA system is conformant with the key standards necessary for implementing security technology with electronic documents, namely XML (Bray et. al., 1998) (with all off-springs like XPath (XPATH, 1999) and XML Schema (XMLSchema, 2001), XSLT (XSLT, 2002)) for representation of e-documents, XMLDSIG (XMLDSIG, 2000) for the format of digital signatures), X.509 (ITU-T, 2000) and the profile defined by IETF in (Housley et. at., 2002) for the format of digital certificates, SSL (SSL, 1996) for the protocol used to provide privacy and reliability between the communicating parties and TSP (Adams et. al., 2001) for the protocol used to support assertions of proof that a datum existed before a particular time. A detailed description of the AIDA architecture and the workflows implemented can be found in (Berbecaru et. al., 2001, 2002).

This paper is focused only on the representation of e-docs in AIDA. The paper is structured as follows: Section 2 explains the main types of cryptographic envelopes used to group altogether the data to be signed, the digital signatures applied on the data and the signer's digital certificate, Section 3 presents the formats of e-docs in AIDA and Section 4 explains how to define the content of an example e-doc that expresses a tax declaration. For simplicity, we'll avoid the complete description of all AIDA components and their functionality. We mention shortly that the overall AIDA architecture is a three-tier architecture composed of: the client side, the server side and the database backend. We'll keep in mind that the e-docs are stored on the database side in a component called *definitions repository* and a dedicated tool, acting at the client side and named *Definitions Manager*, is used for creating the data structures that describe the content of e-docs for a specific workflow. These data structures are called document type definitions and, because are digitally signed, are e-docs themselves.

## 2. DIGITAL SIGNATURE FORMATS

### 2.1 Definition of digital signature

A *digital signature* is a mark that only the sender can make, but other people can easily recognize as belonging to the sender (Pfleeger, 1997). A digital signature is used to confirm agreement to a message, just like a real signature does. A digital signature must meet two primary conditions:
1. *unforgeable*. If sender Alice signs message M with signature S(Alice,M), it is impossible for anyone else to produce the pair [M,S(Alice,M)].
2. *authentic*. If a receiver Bob receives the pair [M,S(Alice,M)], purportedly from Alice, Bob can check that the signature is really from Alice. Only Alice could have created this signature, and the signature is firmly attached to M.
In a public key cryptosystem, Message (M) is signed by appending to it an enciphered summary of the information. The summary is produced by means of a one-way hash function h(), while the enciphering is carried out using an encryption function Encrypt() and the secret key of the signer. Thus, the digital signature performed by Alice on the message M is expressed as:

$$S(Alice,M) = Encrypt(PvK_{Alice}, h(M))$$

The encipherment using the private key ensures that the signature cannot be forged. The one-way nature of the hash function ensures that false information, generated so as to have the same hash result (and thus signature), cannot be substituted. The receiver Bob gets the following signed information [M, S(Alice,M)], composed of the initial data M, and the digital signature S(Alice,M). Bob verifies the Alice's signature by applying first the one-way hash function to the initial information, h(M) and comparing afterwards the result from the above point with that obtained by deciphering with the function Decrypt() the signature S(Alice,M) using the public key of the signer,

$$h(M) = Decrypt(S(Alice,M), PbK_{Alice}) = Decrypt(Encrypt(PvK_{Alice}, h(M)), PbK_{Alice})$$

If the two parts match then Bob can conclude that S(Alice,M) is a valid signature is for the message M. We must note here that if there is not a match between the result of the first step, h(M) and the result of deciphering the signature using the public key of the signer (Alice), then what the receiver (Bob) gets is an error message. Bob cannot determine precisely at this point whether the message (M) was modified, the signature S(Alice,M) was modified or whether it was not the intended person that made the signature (Alice).

The cryptographic envelopes were defined to express the relation among the data to be signed and the digital signature. In this sense we can distinguish: *enveloping* signatures when the data content is inserted in the digital signature envelope; *enveloped* signatures when the signature is inserted in the data content that need to be signed; *detached* signatures when the digital signature object is completely separated from the data content that need to be signed.

For the *detached* signatures it is not specified any way to establish a strict correlation between the data that was signed and the digital signature object. For this reason it is required to have an external mechanism to maintain the correspondence between them, like for example to store them in the same database and to link them in some way. This kind of problem is solved in the *enveloping* signature. The digital signature is bound to the data by mean of some syntax that specifies the format of digital signature like for example PKCS#7/CMS (Housley, 1999). Any number of signers in parallel uses the *signed-data* content type when it is the need to add a digital signature to arbitrary data content. Format for data to be signed is blob (block of bits). The syntax supports also *detached* signatures that is the data and the corresponding signatures are not grouped together in a whole message of *signed-data* content type; in this case the mechanism for the verification of the signatures is application dependent. Instead, in *enveloped* signatures, the digital signature is part of the document itself. For example, Adobe Acrobat supports signatures embedded into data in Portable Document Format (PDF).

### 2.2 XML Signature

XML (eXtensible Markup Language) is a meta-language, i.e. a language that allows defining other languages. XML is part of a more complex language named Standard Generalized Markup Language (SGML) defined in the ISO standard (8879:1986). The language defined for expressing digital signatures is XML Signature. The most important element defined in the XML Signature standard is the *Signature* element.

Since XML is used in AIDA to express the e-doc we will shortly sketch the main parts of the *Signature* element here. As known from the paragraph above on PKCS#7/CMS, the hash value of the data to be signed is not actually signed directly. Rather the hash values of all data elements that should be signed are listed, and then the hash value of this list is signed. It is a kind of double indirection. The compulsory *SignedInfo* element (lines 02 to 10 in Table 1) holds this list of references that need to be signed.

```
[01]<Signature Id="ExampleXMLSignature"
xmlns="http://www.w3.org/2000/07/xmldsig#">
[02]<SignedInfo >
[03]<CanonicalizationMethod Algorithm= …/>
[04]<SignatureMethod Algorithm=
"http://www.w3.org/2000/09/xmldsig#rsa-sha1"/>
[05]<Reference URI="http://www.w3.org/…">
[06]<Transforms> <Transform …/> </Transforms>
[07] <DigestMethod Algorithm=
"http://www.w3.org/2000/07/xmldsig#sha1"/>
[08] <DigestValue> j6lwx3rv=…< /DigestValue >
[09]</Reference>
[10]</SignedInfo>
[11]<SignatureValue> MC=...< /SignatureValue >
[12]<KeyInfo>
[13] <KeyValue> … </ KeyValue>
[14]</KeyInfo>
[15]</Signature>
```

Table 1: Example of Signature element

Each reference stored in the *Reference* element holds a reference identifier, which is normally a URI (attribute in the element *Reference*), the method used for the computation of the hash (the element *DigestMethod*) and a hash value of the referenced data (the element *Digest Value*). When we say "reference to an element", this refers to an element with all its XML element attributes and all descendant elements; this is the whole sub-tree with the denoted element as its root.

The *KeyInfo* element (lines 12 to 14 in Table 1) holds the signing certificate or a reference to it.
This element indicates the key that must be used to validate the signature. Possible forms of identification of this data include certificates, key names, and algorithms for the key exchange. Because the *KeyInfo* element is placed outside the *SignerInfo* element, if the signer wants to bind to the signature the information related to the key, it is possible to use a *Reference* element that can easily identify and include *KeyInfo* as part of the signature itself. The validation of the *SignedInfo* consists of two obligatory processes: the validation of the signature applied on the *SignedInfo* structure and the validation of each hash of the *Reference* structures from the *SignedInfo*. The calculated signature value is in the *SignatureValue* element as base-64 encoded binary data. We can note also that the algorithms used in the calculation of the *SignatureValue* are included in the signed data, while the element *SignatureValue* is external to the element *SignedInfo*.

XML Signature can be used to sign entire XML documents or parts of them or any binary data. XML Signature offers support to *enveloped signature* only for XML data, and to *enveloping* and *detached* signatures both for XML documents and for generic blobs.

## 3. PROPOSED FORMAT FOR E-DOCUMENT

We have stated that e-docs are in general signed documents expressed in electronic form. An example of such a widely used signed document format is signed email, which uses the PKCS#7 format to encode its data.

AIDA is closely related to the reference architecture of the Workflow Management Coalition (Hollingsworth, 1995) established to identify the common characteristics and develop appropriate specifications for implementation of workflow products. Figure 1 shows the basic characteristics of workflow management systems and the relationships between these main functions.

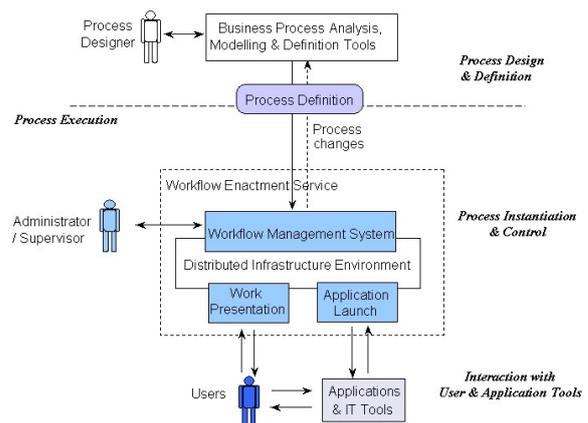

Figure 1. Workflow System Characteristics

At the highest level, all workflow management systems may be characterized as providing support in three functional areas:
1. the build-time functions: these are used for the definition, the modelling and the analysis of workflow processes and related activities,
2. the run-time process control functions: these are concerned with managing the execution of workflow processes,
3. the run-time activity interactions: these enable the cooperation with human users and IT applications during the execution of various steps of workflow processes.
In a document management system, one of the build-time functions is the definition of the format of the e-doc.

In our approach, a signed electronic document consists of the following parts:
1. the signed content;

2. one or more independent parallel signatures, i.e. the signatures applied on the same content. Each signature has signed and unsigned attributes.

We'll describe further how XML was used in AIDA to encode e-docs. A detailed explanation can be found in (Scheilbelhofer, 2001). In AIDA, the e-doc is a well-formed XML document. We choose XML as format to represent an e-doc since this it supports strict separation of content and presentation. Document formats in use today (PDF, Word, HTML) usually mix up data and presentation, making them inappropriate to use when the same information need to be presented differently. Any format mixing presentation and content and that allows active content is also risky to sign: macros, white fonts on white background are only some of the threats. Moreover, some of the formats (e.g. Word's .doc files) have non-public definitions and hence cannot be used in security-critical applications.

The basic structure of an e-doc in AIDA is represented as a tree where the root element of the document is *eDocument*. The first child element is the *signedContent* element, which holds the content data that should be signed. The content can be an *eDocument* once again or it can be any other well-formed XML document. Following the *signedContent*, the root element has one or more *Signature* elements. The structure of the *Signature* element and its children is compliant to the XML Signature format. The *SignedInfo* element of the *Signature* element contains references to:
1. the *signedContent* element, which contains the content data to be signed;
2. the *KeyInfo* element holds the signing certificate or a reference to it;
3. *signedAttributes*.

The *Signature*'s Object element contains *signed* and *unsigned attributes*. The *unsignedAttributes* element can hold any attributes that cannot or should not be covered by the signature value. Normally, this is used for attributes that are not available at the creation time of the signature. Typical unsigned attributes are timestamps, revocation information like CRLs or additional certificates. The structure of the *signedAttributes* element will be defined in the upcoming document of the ETSI for XML Electronic Signatures. This document will define similar structures for XML as already defined for CMS in the ETSI document Electronic Signature Formats (ETSI, 2002).

## 4. EXAMPLE E-DOCUMENT DEFINITION AND MANAGEMENT

This section explains how to define and manage an example e-doc representing a sample tax declaration. For any e-doc it is required to define two pieces of information: the content of the e-doc and how the content is to be displayed. In other words, to implement any workflow that uses tax declaration e-docs, someone (usually the process designer, that is the person in charge with the process design and definition) must define first the document's type definition and the display transformations for that type of e-doc. A type definition contains the e-doc's content while the display transformation contains the XML transforms (or stylesheets) used to present the e-docs content to the user. To create and sign the type definitions and the transformations, the *Definitions Manager* tool is used. In AIDA, all type definitions and transforms stored in the *definitions repository* must have a unique ID, which is assigned at the definition time. A complete type definition contains a document type ID expressed by the *<aida:documentTypeID>* element and the actual XML Schema contained in the element *<aida:schema>*. The document type ID is up to 100 characters long and looks like a URL, such as "aida://www.polito.it/tax". For the definition of the content of an e-doc, the AIDA system allows to define either a *generic* type definition or a *specific* type definition.

A *generic* type definition is characterized by its simplicity. Anyone can easily create such a generic schema by defining only the name and types of the fields, without knowing other details of XML Schema. The generic schema is expressed by the element *<aida:genericSchema>*. A type definition of this kind is processed further with the *Definitions Manager* tool, which will ask the process designer for the preferred document type ID. The tool will create automatically the XML Schema and the enveloping AIDA e-doc. A *specific* type definition instead can be any correct XML Schema definition. The generic type definition for an example e-doc expressing a tax declaration is shown in Table 2.

```
<aida:genericSchema
xmlns:aida="http://aida.infonova.at">
<aida:documentRoot> tax
</aida:documentRoot>
<aida:documentNamespace>
http://www.polito.it/tax
</aida:documentNamespace>
<aida:namespacePrefix> polito
</aida:namespacePrefix>
<aida:fieldList>
<aida:field>
    <aida:name> Unique_identification_number
    </aida:name>
<aida:shortString max="20"
searchable="true"/>
</aida:field>
<aida:field>
    <aida:name>Surname</aida:name>
  <aida:shortString max="20"
searchable="true"/>
</aida:field>
<aida:field>
    <aida:name>Name</aida:name>
  <aida:shortString max="20"
searchable="true"/>
</aida:field>
<aida:field>
    <aida:name>Income_from_buildings_fields
    </aida:name>
    <aida:shortString max="70"/>
</aida:field>
<aida:field>
    <aida:name>Income_as_employee
    </aida:name>
    <aida:shortString max="70"/>
</aida:field>
<aida:field>
```

```xml
    <aida:name>Other_incomes</aida:name>
    <aida:shortString max="80"/>
</aida:field>
<aida:field>
    <aida:name>Taxes_Expenses</aida:name>
    <aida:shortString max="80"/>
</aida:field>
<aida:field>
    <aida:name>Phone_number</aida:name>
    <aida:shortString max="30"/>
</aida:field>
<aida:field>
    <aida:name>Mail_address</aida:name>
    <aida:shortString max="200"/>
</aida:field>
</aida:fieldList>
</aida:genericSchema>
```

Table 2. Generic type definition for tax declaration e-doc

The generic type definition contains:
1. the name of the e-doc XML root element contained in the tag *<aida:documentRoot>*
2. the XML namespace identifier contained in the element *<aida:documentNamespace>*
3. the XML namespace prefix to be used for each field element contained in the element *<aida:namespacePrefix>*
4. the actual list of fields of the e-doc

All the XML tags are prefixed with the *aida:* namespace prefix and the *xmlns:aida="http://aida.infonova.at"* XML namespace identifier is used in the root element. The *<aida:fieldList>* tag can contain an arbitrary number of *<aida:field>* tags. Each *<aida:field>* tag must contain one *<aida:name>* tag with the desired field name and one empty tag denoting the field type. The field types can take one of the values shown in Table 3.

| Name | XML Schema datatype | Comment |
|---|---|---|
| aida:string | string | no length limit |
| aida:shortString | string | maximum 250 characters |
| aida:date | date | fixed format YYYY-MM-DD |
| aida:time | time | fixed format HH:MM:SS.SSS |
| aida:int | int | signed 32-bit integer in canonical textual form |
| aida:double | double | IEEE standard double 64-bit in canonical textual form |
| aida:Boolean | boolean | "true" or "false" |

Table 3. Simple AIDA field types

Using the *generic* type definition for the example tax declaration e-doc, the process designer can build the document type definition for the tax declaration having *<aida:documentTypeData>* as root element. The AIDA type definition can be further exported in an XML file having the content illustrated in the Table 4. Such document type definitions are constructed from generic or specific type definition input files with the *Definitions Manager* tool but they can also be constructed by hand since the structure of the enveloping *<aida:documentTypeData>* element is not complex.

```xml
<?xml version="1.0" encoding="UTF-8" ?>
<aida:documentTypeData
xmlns:aida="http://aida.infonova.at"
xmlns:xsi=http://www.w3.org/2001/XMLSchema-instance
xsi:schemaLocation="http://aida.infonova.at
aida:documentTypeData">
<aida:documentTypeID>
    aida://www.polito.it/tax
</aida:documentTypeID>
<aida:schema>
  <xsd:schema
targetNamespace="http://www.polito.it/tax"
xmlns:polito="http://www.polito.it/tax"
xmlns:xsd="http://www.w3.org/2001/XMLSchema">
<xsd:element name="tax">
 <xsd:complexType>
    <xsd:sequence>
      <xsd:element
ref="polito:Unique_identification_number"/>
      <xsd:element ref="polito:Surname"/>
      <xsd:element ref="polito:Name"/>
      <xsd:element
ref="polito:Income_from_buildings_fields"/>
      <xsd:element
ref="polito:Income_as_employee"/>
      <xsd:element ref="polito:Other_incomes"/>
      <xsd:element ref="polito:Taxes_Expenses"/>
      <xsd:element ref="polito:Phone_number"/>
      <xsd:element ref="polito:Mail_address"/>
    </xsd:sequence>
 </xsd:complexType>
</xsd:element>
<xsd:simpleType name="shortString">
  <xsd:restriction base="xsd:string">
    <xsd:maxLength value="250"/>
  </xsd:restriction>
</xsd:simpleType>
<xsd:element
name="Unique_identification_number">
  <xsd:simpleType>
    <xsd:restriction
base="polito:shortString">
      <xsd:maxLength value="20"/>
    </xsd:restriction>
  </xsd:simpleType>
</xsd:element>
  ...
 </xsd:schema>
</aida:schema>
<aida:genericSchema
xmlns:aida="http://www.polito.it">
… data from Table 2
 </aida:genericSchema>
</aida:documentTypeData>
```

Table 4. Document type definition built from generic schema

To be stored in the definitions repository, the document type definition is further signed with the *Definitions Manager*. The e-doc containing the document type definition that has been digitally signed is illustrated in Table 5. We can note that the signed document type definition is an e-doc itself as it is contained in the element *<aida:eDocument>* and has the format described in Section 3. The signature is applied on the element *<aida:signedContent>* that contains at its turn the element *<aida:documentTypeData>* and its child elements that were shown in Table 4. The XML signature element is contained in the element *<dsig:Signature>*.

```xml
<?xml version="1.0" encoding="UTF-8"?>
```

```xml
<aida:eDocument
xmlns:aida="http://www.polito.it"
xmlns:xsi=http://www.w3.org/2001/XMLSchema-instance
xsi:schemaLocation="http://www.polito.it
aida:eDocument">
<aida:signedContent>
<aida:documentTypeData
… data from Table 4
</aida:documentTypeData>
</aida:signedContent>
<dsig:Signature
xmlns:dsig="http://www.w3.org/2000/09/xmldsig
#">
<dsig:SignedInfo>
<dsig:CanonicalizationMethod
Algorithm="http://www.w3.org/TR/2001/REC-xml
-c14n-20010315"/>
<dsig:SignatureMethod Algorithm=
"http://www.w3.org/2000/09/xmldsig#rsa-
sha1"/>
<dsig:Reference
URI="#xmlns(aida=http://www.polito.it)… ">
<dsig:DigestMethod
Algorithm="http://www.w3.org/2000/09/xmldsig#
sha1"/>
<dsig:DigestValue> 8/E+Xcs=…
</dsig:DigestValue>
</dsig:Reference>
<dsig:Reference
URI="#xmlns(aida=http://www.polito.it)… ">
<dsig:DigestMethod
Algorithm="http://www.w3.org/2000/09/xmldsig#
sha1"/>
<dsig:DigestValue>1Gz1wk43I=
</dsig:DigestValue>
</dsig:Reference>
<dsig:Reference
URI="#xmlns(aida=http://www.polito.it)… ">
<dsig:DigestMethod Algorithm=
"http://www.w3.org/2000/09/xmldsig#sha1"/>
<dsig:DigestValue>onIoD=..</dsig:DigestValue>
</dsig:Reference>
</dsig:SignedInfo>
<dsig:SignatureValue> JZLLunRKQFxaw+GlS…
</dsig:SignatureValue>
<dsig:KeyInfo>  <dsig:X509Data>
   <dsig:X509Certificate>IBATAJBg…
   </dsig:X509Certificate>
   </dsig:X509Data>  </dsig:KeyInfo>
<dsig:Object>
  <aida:properties
xmlns:aida="http://www.polito.it">
   <aida:signedProperties/>
   <aida:unsignedProperties/>
  </aida:properties>
</dsig:Object>
</dsig:Signature>
</aida:eDocument>
```

Table 5. Signed document type definition for tax declaration e-doc

AIDA makes use of XML transforms (also called stylesheets) to present the e-doc's content to the user. In general, the main purpose of the transforms is to separate the presentation of data from the content. For the same e-doc in XML format one or more XSL stylesheet can be written to control the output of the e-doc to the screen, printed page, or any other two dimensional display devices. For the tax declaration e-doc, the XSL stylesheet used to render the content is illustrated in Table 6. The output format created by this stylesheet is "MHTML", a subset of HTML that can be processed by the client modules.

```xml
<xsl:stylesheet
xmlns:xsl="http://www.w3.org/1999/XSL/Transf
orm" version="1.0"
xmlns:polito="http://www.polito.it/tax">
<xsl:output method="xml"/>
<xsl:template match="polito:tax">
<aida:mhtml
xmlns:aida="http://www.polito.it"
xmlns:xsi="http://www.w3.org/2001/XMLSchema-
instance"
xsi:schemaLocation="http://www.polito.it
aida:mhtml">
<aida:body>
<aida:b>
<aida:font size="+2">Tax Declaration
</aida:font>
</aida:b><aida:br/><aida:br/>
<aida:br/>
<aida:font size="+0">Fiscal Code:
</aida:font>
<aida:b><xsl:value-of
select="polito:Unique_identification_number"
/>
</aida:b>
<aida:br/><aida:font size="+0">Surname:
</aida:font>
<aida:b><xsl:value-of
select="polito:Surname"/>
</aida:b>
<aida:br/><aida:font size="+0">Name:
</aida:font>
<aida:b><xsl:value-of select="polito:Name"/>
</aida:b>
<aida:br/>
<aida:font size="+0">
Income from buildings and fields:
</aida:font>
<aida:b><xsl:value-of
select="polito:Income_from_buildings_fields"
/>
</aida:b>
<aida:br/><aida:font size="+0">Income as
employee: </aida:font>
<aida:b><xsl:value-of
select="polito:Income_as_employee"/>
</aida:b>
<aida:br/><aida:font size="+0">Other
incomes: </aida:font>
<aida:b><xsl:value-of
select="polito:Other_incomes"/></aida:b>
<aida:br/><aida:font size="+0">Taxes
Expenses: </aida:font>
<aida:b><xsl:value-of
select="polito:Taxes_Expenses"/> </aida:b>
<aida:br/><aida:font size="+0">Phone Number:
</aida:font>
<aida:b><xsl:value-of
select="polito:Phone_number"/> </aida:b>
<aida:br/><aida:font size="+0">Mail Address:
</aida:font>
<aida:b><xsl:value-of
select="polito:Mail_address"/>
</aida:b>
</aida:body></aida:mhtml>
</xsl:template>
</xsl:stylesheet>
```

Table 6. XSL stylesheet for tax declaration e-doc

From the stylesheet shown in the Table 6, the process designer will build and sign a transform with the Definitions Manager. At this step the process definer must indicate:
1. a transform identifier, whose value will be contained in the element *<aida:transformDataID>*
2. the e-document type identifier for which the transform is intended, whose value will be contained in the element *<aida:documentTypeID>*
3. the transform method, contained in the element *<aida:transformMethod>*
4. language, contained in the element *<aida:language>*

5. the output format, contained in the element *<aida:outputFormat>*

The resulting transform will be an e-doc itself, as illustrated in Table 7. We can note the element *<aida:eDocument>* which marks the beginning of an e-doc, the signed content is the data contained in the tree *<aida:signedContent>*, and the signature applied on the transform is contained in the *<dsig:Signature>* element.

```
<?xml version="1.0" encoding="UTF-8"?>
<aida:eDocument
xmlns:aida="http://www.polito.it"
xmlns:xsi="http://www.w3.org/2001/XMLSchema-
instance"
xsi:schemaLocation="http://www.polito.it
aida:eDocument">
<aida:signedContent>
<aida:transformData
xmlns:aida="http://www.polito.it"
xmlns:xsi="http://www.w3.org/2001/XMLSchema-
instance"
xsi:schemaLocation="http://www.polito.it
aida:displayData">
<aida:transformDataID>taxTrafo1
</aida:transformDataID>
<aida:documentTypeID>aida://www.polito.it/tax
</aida:documentTypeID>
<aida:requiredDisplayCapabilities>
<aida:transformMethod>xslt
</aida:transformMethod>
<aida:language>en</aida:language>
<aida:outputFormat>mhtml</aida:outputFormat>
</aida:requiredDisplayCapabilities>
<aida:transform>
<aida:documentFrameXSLStylesheet>
  <xsl:stylesheet …
    …..data from Table 6
  </xsl:stylesheet>
</aida:documentFrameXSLStylesheet>
</aida:transform>
</aida:transformData>
</aida:signedContent>
<dsig:Signature
xmlns:dsig="http://www.w3.org/2000/09/xmldsig
#">
<dsig:SignedInfo>
<dsig:CanonicalizationMethod
Algorithm="http://www.w3.org/TR/2001/REC-xml-
c14n-20010315"/>
<dsig:SignatureMethod
Algorithm="http://www.w3.org/2000/09/xmldsig#
rsa-sha1"/>
<dsig:Reference … reference toward
aida:signedContent of aida:eDocument ">
<dsig:DigestMethod
Algorithm="http://www.w3.org/2000/09/xmldsig#
sha1"/>
<dsig:DigestValue>zzyv8=…</dsig:DigestValue>
</dsig:Reference>
<dsig:Reference  … reference toward
dsig:KeyInfo element of dsig:Signature
element ">
<dsig:DigestMethod
Algorithm="http://www.w3.org/2000/09/xmldsig#
sha1"/>
<dsig:DigestValue>94W1G=…</dsig:DigestValue>
</dsig:Reference>
<dsig:Reference … reference toward aida:
signedProperties of the dsig:Object element">
<dsig:DigestMethod
Algorithm="http://www.w3.org/2000/09/xmldsig#
sha1"/>
<dsig:DigestValue>onIow=…</dsig:DigestValue>
</dsig:Reference>
</dsig:SignedInfo>
<dsig:SignatureValue>2nLmaWWfA=…
</dsig:SignatureValue>
<dsig:KeyInfo>
  <dsig:X509Data>
    <dsig:X509Certificate>CAjegAwIBAgIB…
    </dsig:X509Certificate>
  </dsig:X509Data>
</dsig:KeyInfo><dsig:Object>
<aida:properties
xmlns:aida="http://www.polito.it">
<aida:signedProperties/>
<aida:unsignedProperties/>
</aida:properties></dsig:Object>
</dsig:Signature>
</aida:eDocument>
```
Table 7. Signed transform for tax declaration e-doc

Once the content and the display information were defined for the e-doc, the task of the process definer is finished. Instances for the e-docs must be further generated as part of the workflow. For example, there exists a component tool in AIDA, called *Scenario Application*, whose task is to model a workflow. Among the *Scenario Application*'s tasks, the primary one is to generate e-doc instances. The *Scenario Application* needs to receive as input the values for the fields (such as the value of the element called ``Surname'') and to generate as output an e-doc instance. An example tax declaration e-doc instance is illustrated in Table 8. Any e-doc instance must be structured in accordance to the content previously defined (shown in Table 4), signed (shown in Table 5) and finally stored in the *definitions repository*. The tools running at the client side must check if the e-doc handled has the structure specified in the previously defined content (in our case the content shown in Table 4). For this purpose, the client tools need to retrieve firstly from the *definitions repository* the signed document type definition (shown in Table 5). At a second step the client tool verifies the integrity of the signed document type definition by verifying the digital signature applied on it. After that, the client tool can check that the e-doc instance (the part of data between the *<aida:signedContent>* and *</aida:signedContent>* in Table 8) respects the structure defined in the document type definition (data between *<aida:schema>* and *</aida:schema>* in Table 5). Finally the client tool checks the integrity of the e-doc instance by checking the digital signature applied on it. For this purpose it uses the part of data contained in the element *<dsig:Signature>* for the information contained in the element *<aida:signedContent>* in Table 8.

```
<?xml version="1.0" encoding="UTF-8"?>
<aida:eDocument xmlns:aida=
http://www.polito.it xmlns:xsi=
http://www.w3.org/2001/XMLSchema-instance
xsi:schemaLocation= "http://www.polito.it
aida:eDocument">
<aida:signedContent>
   <polito:tax
xmlns:polito="http://www.polito.it/tax"
xmlns:xsi=http://www.w3.org/2001/XMLSchema-
instance
xsi:schemaLocation="http://www.polito.it/tax
aida://www.polito.it/tax">
<polito:Unique_identification_number>D12876
   </polito:Unique_identification_number>
   <polito:Surname>Popescu</polito:Surname>
   <polito:Name>Ion</polito:Name>
   <polito:Income_from_buildings_fields>
     21,000,000
   </polito:Income_from_buildings_fields>
   <polito:Income_as_employee> 2,000,000
   </polito:Income_as_employee>
```

```xml
    <polito:Other_incomes> 1,000,000
    </polito:Other_incomes>
    <polito:Taxes_Expenses> 2,000,000
    </polito:Taxes_Expenses>
    <polito:Phone_number> +22323214
    </polito:Phone_number>
    <polito:Mail_address>
Popescu.Ion@domain.com
    </polito:Mail_address>
  </polito:tax>
</aida:signedContent>
<dsig:Signature
xmlns:dsig="http://www.w3.org/2000/09/xmldsig
#">
<dsig:SignedInfo>
  <dsig:CanonicalizationMethod
Algorithm="http://www.w3.org/TR/2001/REC-xml-
c14n-20010315"/>
<dsig:SignatureMethod Algorithm=
"http://www.w3.org/2000/09/xmldsig#rsa-
sha1"/>
<dsig:Reference
URI="#xmlns(aida=http://www.polito.it)… ">
<dsig:DigestMethod Algorithm=
"http://www.w3.org/2000/09/xmldsig#sha1"/>
<dsig:DigestValue>k+nbo…=</dsig:DigestValue>
</dsig:Reference>
<dsig:Reference URI=
"#xmlns(aida=http://www.polito.it)… ">
<dsig:DigestMethod Algorithm=
"http://www.w3.org/2000/09/xmldsig#sha1"/>
<dsig:DigestValue>J1KMyLX=…/dsig:DigestValue>
</dsig:Reference>
<dsig:Reference URI=
"#xmlns(aida=http://www.polito.it)… ">
<dsig:DigestMethod Algorithm=
"http://www.w3.org/2000/09/xmldsig#sha1"/>
<dsig:DigestValue>hD/HNk=…</dsig:DigestValue>
 </dsig:Reference>
</dsig:SignedInfo>
<dsig:SignatureValue>NtfU6BymvR…
</dsig:SignatureValue>
<dsig:KeyInfo>
   <dsig:X509Data>
     <dsig:X509Certificate> 8TCCAl6gAwIB…
     </dsig:X509Certificate>
   </dsig:X509Data> </dsig:KeyInfo>
<dsig:Object>
 <aida:properties
xmlns:aida="http://www.polito.it">
   <aida:signedProperties>
     <aida:transformDataID>taxTrafo1
     </aida:transformDataID>
     <aida:documentHash>
       tVzJfpiBHZrJN=…
     </aida:documentHash>
   </aida:signedProperties>
   <aida:unsignedProperties>
     <aida:signatureValueTimeStamp>
       MIIULjADAgEAMIIUJQYJK…
     </aida:signatureValueTimeStamp>
   </aida:unsignedProperties>
 </aida:properties>
</dsig:Object>
</dsig:Signature>
</aida:eDocument>
```

Table 8. E-document instance for tax declaration

## 5. CONCLUSIONS

AIDA Project implemented a secure XML-based document management system that can be easily adapted to different types of real world scenarios where e-docs must be digitally signed. One of the main requirements in the design phase was to analyse and choose a format for the digital signature and to design and implement the format of the e-doc. This paper describes the most widely used formats for signature cryptographic envelopes and describes the representation of e-doc in AIDA based on XML and XML Signature.

**Acknowledgements** – Work carried out under the financial support of European Commission under project AIDA (IST-1999-10497).